\documentclass[conference,10pt]{IEEEtran}
\IEEEoverridecommandlockouts
\usepackage{cite}
\usepackage{amsmath,amssymb,amsfonts}
\usepackage{algorithmic}
\usepackage{graphicx}
\usepackage{textcomp}
\usepackage{xcolor}
\usepackage{subfig}
\graphicspath{{figures/drawings/}{figures/plots/pdf/}}
\def\BibTeX{{\rm B\kern-.05em{\sc i\kern-.025em b}\kern-.08em T\kern-.1667em\lower.7ex\hbox{E}\kern-.125emX}}
\begin{document}

\title{ Distributed Transmit Beamforming: Analyzing the Maximum Communication Range
}

\author{\IEEEauthorblockN{Samer Hanna\IEEEauthorrefmark{1} and Danijela Cabric\IEEEauthorrefmark{1}\thanks{This work was supported in part by NSF  under  grant 1929874 and by the CONIX Research Center, one of six centers in JUMP, a Semiconductor Research Corporation (SRC) program sponsored by DARPA.}}\IEEEauthorblockA{\IEEEauthorrefmark{1}Electrical and Computer Engineering Department, University of California, Los Angeles}
\IEEEauthorblockA{samerhanna@ucla.edu, danijela@ee.ucla.edu}}

\newtheorem{prop}{Proposition}
\maketitle

\begin{abstract}
Distributed transmit beamforming is a technique that adjusts the signals from cooperating radios to combine  coherently at a destination radio. To achieve coherent combining, the radios can exchange preambles with the destination for frequency synchronization and signal phase adjustment. At the destination, coherent combining leads to a  beamforming (BF) gain. The BF gain can extend the communication range by countering the path loss that increases with the distance from the destination. While ideally the maximum range can be trivially  calculated from the BF gain, in reality, the BF gain depends on the distance because, at a larger distance, lower SNR of the exchanged preambles causes higher synchronization and phase estimation errors, which in turn degrades the BF gain.  In this paper, considering the BF gain degradation for a destination-led BF protocol, we calculate the maximum communication range  to realize a desired post-BF SNR by analyzing the relation between the pre-BF SNR and the BF gain. We show that  increasing the  preamble lengths or increasing the destination power can significantly increase the maximum range  while just increasing the number of radios gives diminishing range extension. 
\end{abstract}


\section{Introduction}

Distributed transmit beamforming (DBF) can be realized by adjusting signals transmitted  from independent cooperative radios such that they combine coherently at a destination radio, as if the signals were transmitted from a virtual antenna array~\cite{mudumbai_feasibility_2007}. For $N$ equal transmit power radios, DBF can increase the received power at the destination by up to $N^2$; $N$  due to the increase in the  transmitted power and $N$  due to the improved directivity leading to a coherent beamforming (BF) gain at the destination~\cite{mudumbai_distributed_2009}. In a line-of-sight (LOS) channel having a path loss exponent  $k=2$, this $N^2$ gain can increase the communication range by up to  $N$  folds ($N^\frac{2}{k}$ in general) while retaining the  power level at the destination. This improvement in communication range can be crucial for remote radio deployments on UAVs~\cite{mohanti_airbeam_2019} or ground vehicles~\cite{kramarev_event-triggered_2019} for applications like surveillance or disaster management.

To achieve coherent combining using DBF in a narrowband channel, the DBF radios having independent oscillators need to first synchronize their carrier frequencies, which can be performed with the assistance of the destination.  Then they need to adjust the  phases of the transmitted signal to compensate for  the channel with the destination. Since the phase adjustment depends on the destination, it either requires a feedback signal from the destination or knowing the DBF radios location along with the BF direction to reach the destination~\cite{jayaprakasam_distributed_2017}.  Feedback based DBF suffers from a low SNR  feedback signal for remote deployments. However, at large distances in a non LOS channel, due to shadowing,  the BF direction to reach the destination can be hard to know making feedback-based DBF more practical method despite  the noisy feedback. The noisy feedback, however, will lead to degraded synchronization leading to a random  BF gain below the ideal $N^2$ and a range extension below the ideal $N^\frac{2}{k}$, which makes designing a DBF system challenging. 
To design a DBF system, for a given number of cooperative radios,  we need to specify the maximum communication range such that a required SNR is met with a given probability at the destination. This is challenging because as the DBF radios get further from the destination, more BF gain is needed to counter the increasing path loss, however,  the SNR of signals exchanged with the destination gets lower making the BF gains  degrade. 

Existing works have proposed and demonstrated many distributed BF approaches. Location based BF towards a known direction in a LOS channel was demonstrated in~\cite{ellison_multi-node_2021}. DBF  was demonstrated using  explicit channel phase  feedback in~\cite{leak_distributed_2018,kramarev_event-triggered_2019,mohanti_airbeam_2019}.  Other approaches used 1-bit feedback to iteratively adjust the phase~\cite{quitin_demonstrating_2012,quitin_scalable_2013,mohanti_wifed_2021}. Estimating the phase in a retrodirective manner relying on reciprocity was also proposed in~\cite{peiffer_experimental_2016}. However, reciprocity does not always apply and iterative approaches are not suitable for fast varying channels, making explicit channel feedback the most reliable approach. While these works have demonstrated many interesting DBF approaches, they lack the analysis  to calculate the maximum communications range.
In our prior work~\cite{hanna_distributed_2021}, for a destination-led DBF protocol using Kalman filter for frequency synchronization and explicit channel estimation and feedback, we analyzed the relation between the preBF SNR and the BF gains. These relations were verified using  simulations and experimentally using software-defined radios.

In this paper, using the DBF protocol and analytical framework from~\cite{hanna_distributed_2021}, we calculate the maximum communication range  that maintains a required  postBF SNR  with a given probability. We study the impact of changing the number of BF radios, the preambles lengths, and the destination transmit power on the maximum communication range.  Our results show that only increasing $N$ might not be the best strategy to extend the communication range.

\newcommand{\mhl}[1]{#1}

\renewcommand{\b}[1]{\boldsymbol{\mathrm{#1}}}
\providecommand{\h}[1]{\ensuremath{\b{h}_{#1}}}

\newcommand{\C}[1]{\mathbb{C}^{#1}}
\newcommand{\R}[1]{\mathbb{R}^{#1}}
\newcommand{\Z}{\mathbb{Z}}
\newcommand{\U}{\mathcal{U}}
\newcommand{\mI}{\b{I}}
\newcommand{\E}[1]{\mathbb{E}\{#1\}}

\newcommand{\mfh}[1]{ f_{#1}}
\newcommand{\mph}[1]{\phi_{#1}}
\newcommand{\mah}[1]{a_{#1}}

\newcommand{\mpwd}[1]{\hat{\phi}'_{#1}}
\newcommand{\mpw}[1]{\hat{\phi}_{#1}}
\newcommand{\mfw}[1]{\hat{f}_{#1}}

\newcommand{\mSNRpre}{\gamma_{\text{preBF}}}
\newcommand{\mSNRpost}{\gamma_{\text{postBF}}}
\newcommand{\mSNRms}{\gamma_{\text{DR}}}
\newcommand{\mSNRmin}{\gamma_{\text{min}}}

\newcommand{\mSNRpreDB}{\Gamma_{\text{preBF}}}
\newcommand{\mSNRpostDB}{\Gamma_{\text{postBF}}}
\newcommand{\mSNRmsDB}{\Gamma_{\text{DR}}}
\newcommand{\mSNRminDB}{\Gamma_{\text{min}}}

\newcommand{\mtsyn}{t_{\text{syn}}}
\newcommand{\mtph}{t_{\text{ph}}}
\newcommand{\mtfb}{t_{\text{fb}}}
\newcommand{\mtgI}{t_{\text{g1}}}
\newcommand{\mtgII}{t_{\text{g2}}}
\newcommand{\mtgIII}{t_{\text{g3}}}
\newcommand{\mtov}{t_{\text{ov}}}
\newcommand{\mtp}{t_{\text{p}}}

\newcommand{\mnsyn}{N_{\text{syn}}}
\newcommand{\mnph}{N_{\text{ph}}}
\newcommand{\mnfb}{N_{\text{fb}}}
\newcommand{\mngI}{N_{\text{g1}}}
\newcommand{\mngII}{N_{\text{g2}}}
\newcommand{\mngIII}{N_{\text{g3}}}
\newcommand{\mlzc}{M}
\newcommand{\mnzc}{N_{\text{ZC}}}
\newcommand{\mnov}{N_{\text{ov}}}

\newcommand{\mte}{t_e}
\newcommand{\mpe}[1]{\phi^e_{#1}}

\newcommand{\mTs}{T_s}

\newcommand{\mvar}[1]{\text{var}\{#1\}}
\newcommand{\mcov}[1]{\text{cov}\{#1\}}
\newcommand{\mvarN}{\sigma^2}
\newcommand{\mvarPe}{\sigma^2_{e}}
\newcommand{\mstdPe}{\sigma_{e}}
\newcommand{\mvarF}{\sigma^2_{f}}
\newcommand{\mvarPh}{\sigma^2_{ph}}
\newcommand{\mvarPhe}{\sigma^2_{phe}}
\newcommand{\mvarFb}{\sigma^2_{fb}}
\newcommand{\mvarFbe}{\sigma^2_{fbe}}
\newcommand{\mPdt}{P^{D}_{T}}

\newcommand{\mfEst}{\eta_f}

\newcommand{\myf}{y_f}
\newcommand{\mxf}{x_f}
\newcommand{\mvarFe}{\sigma^2_{fe}}
\newcommand{\mvarFk}{\sigma^2_{fk}}
\newcommand{\mtcycle}{t_{\text{cyc}}}

\newcommand{\mxcf}{X_f}
\newcommand{\mxcp}{X_p}
\newcommand{\mstdcf}{\sigma_2}
\newcommand{\mstdcp}{\sigma_1}
\newcommand{\mwinf}{W_2}
\newcommand{\mwinp}{W_1}
\newcommand{\mcoht}{\tau_c}
\newcommand{\mcohl}{R_L}

\newcommand{\mGreq}{G_{\text{req}}}
\newcommand{\mNmin}{N_{\text{lb}}}
\newcommand{\mNmax}{N_{\text{ub}}}

\newcommand{\mpout}{p_{\text{out}}}

\newcommand{\mdp}{\Delta P}
\newcommand{\mGdb}{G_{dB}}

\newcommand{\mSNRreq}{\gamma_{\text{req}}}
\newcommand{\mSNRreqDB}{\Gamma_{\text{req}}}

\newcommand{\mnTrgt}{L} 

\newcommand{\mpmin}{p_{\text{min}}}

\section{System Model and Problem Statement}
We consider $N$ equal transmit power radios, each having power $P_T$, cooperating to transmit a shared payload $m(t)$ to a remote destination $D$ using DBF. The destination radio is assumed to have a transmit power which is $\mdp$ dB higher than $P_T$, where $\mdp\ge 0$dB.
 Each of the DBF radios has an independent oscillator such that radio $n$ has a carrier frequency offset $\mfh{n}$ and a phase offset $\mph{n}$ with respect to the destination. These offsets are assumed to be constant within the BF payload, whose duration is assumed to be less than the channel coherence time. Assuming the DBF radios precode their signals and transmit them simultaneously at their maximum power, the received signal at the destination is given by
$
	\label{eq:rec_sig}
	y(t)=  \sqrt{P_T}/a  \sum_{n=1}^{N}    z_n(t)  \exp \{ j(2\pi \mfh{n} t + \mph{n} )\} + w(t)
$
where  $w(t)$ is the white Gaussian noise process having power spectral density $N_0/2$. The DBF radios are assumed in proximity of each other far from the destination by a distance $d$ and thus experience the same path loss denoted by $a= \frac{\lambda}{2\pi} d^{k}$, with $\lambda$ being the wavelength and $k$ the path loss exponent.  The precoded signal sent by radio $n$ is given by
$
	z_n(t)= m(t)  \exp \{-j(2\pi \mfw{n} t + \mpw{n} )\}
$
where $\mfw{n}$ and $\mpw{n}$ are the $n$-th radio corrections of the frequency and phase offsets respectively, which are estimated through the DBF protocol. These estimates, however, are not perfect, especially at a low SNR, which would result in residual frequency and phase offsets leading to a cumulative phase errors given by 
$
	\label{eq:phase_error}
	\mpe{n}(t) = (2\pi (\mfh{n}-\mfw{n}) t + (\mph{n}-\mpw{n} ) )
$
making the received signal equal to
$
	y(t)=  \frac{\sqrt{P_T}}{a}  m(t) \sum_{n=1}^{N}    \exp \{ j \mpe{n}(t) \} + w(t)
$.
At the evaluation time  $\mte$, the BF gain caused by coherent combining is given by 
$
	\label{eq:bf_gain}
	G=\frac{1} { N }  \left| \sum_{n=1}^{N}  \exp \{ j \mpe{n} \}\right|^2
$ and $NG$ is referred to as the total BF gain, which can take value up to $N^2$.
 However, due to the random phase error, the combining BF gain $G$ becomes a random variable making the total BF gains take values below $N^2$.
 
 The pre-BF SNR at the destination from one radio  is  
$
	\mSNRpre=\frac{ P_T}{a^2 N_0} 
$
and it degrades with the distance $d$, making the maximum range corresponding to the smallest $\mSNRpre$. 
The post-BF SNR of the combined signal from all $N$ BF radios is equal to 
$
	\label{eq:postBF_SNR}
	\mSNRpost = N G \mSNRpre   
$. It is random and follows the same distribution as  $G$.
 According to this relation, in order to maintain the same postBF SNR at the destination, when deploying the DBF radios at a further distance (smaller $\mSNRpre$), requires increasing $G$. However,  the coherent BF gain $G$ depends on the quality of synchronization,  and  degrades with smaller $\mSNRpre$.

In order to specify the largest distance $d$ that realizes a desired postBF SNR  $\mSNRreq$ with a given minimum probability $\mpmin$, we formulate the following optimization problem
\begin{align}
	& \underset{x}{\text{maximize}} 	& & d \label{probI:obj} \\
	& \text{subject to}	& & \mSNRpost = N \mSNRpre(d)   G(\mSNRpre(d))   \label{probI:sum}\\
	& & & P(\mSNRpost \geq \mSNRreq ) \geq \mpmin \label{probI:probability}
\end{align}
To solve this problem, we  first obtain an analytical relation between the BF gain $G$ and the preBF SNR for a given BF protocol. Then, we derive the distribution of $\mSNRpost$, caused by the randomness in the coherent combining gain $G$.

\section{DBF Protocol and Coherent BF Gain}
\begin{figure}[t!]
	\centerline{\includegraphics[scale = 0.95]{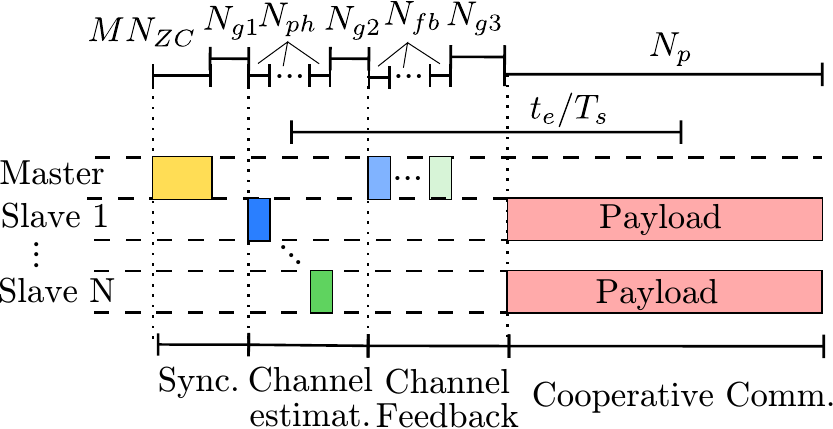}}
	\caption{The timing diagram of the DBF protocol.}
	\label{fig:timing_diagram}
\end{figure}
\subsection{Protocol}
We consider the master-slave beamforming protocol from~\cite{hanna_distributed_2021} ; the destination is the master and the DBF radios are the slaves.  The master initiates the beamforming procedure and sends a synchronization (synch) preamble for time alignment and frequency offset estimation.  After correcting their frequencies, the slaves send channel estimation preambles to the master, each in a predetermined time-slot. The master  calculates a phase estimate $\mpwd{n}$  and transmits it back to each slave in a predetermined time-slot. The slaves receive a slightly different value $\mpw{n}$ due to feedback errors. Once each slave knows $\mpw{n}$ and $\mfw{n}$, they  start transmitting their payload.  

In Fig.~\ref{fig:timing_diagram}, we illustrate the transmitted signals, where their duration is expressed in discrete time assuming a sampling time $T_s$. The synchronization (synch) preamble  consists of $\mnzc$ repetition of a Zadoff-Chu (ZC) sequence of length $\mlzc$ similar to~\cite{yan_aeroconf_2019}, using $\mnzc \mlzc$ samples. For phase estimation, each slave transmits a known signal  consisting of $\mnph$ samples. For phase feedback of slave $n$, the value of $\mpwd{n}$  is encoded in the phase difference between two identical preambles of length $\mnfb$. Guard times are allocated after each stage for processing ($\mngI,\mngII,\mngIII$) and the payload consists of $N_p$ samples.
 All the signaling is performed on the same frequency band, hence, all transmissions are received by all radios.  The duration of the BF   overheads incurred by the protocol in samples is given by
$
	\label{eq:overhead_n}
	\mnov = \mnsyn + N (\mnph + \mnfb) + \mngI +\mngII + \mngIII 
$.
Note the channel estimation and feedback is performed per slave, and hence the larger $N$ the shorter the preambles have to be for the overhead to remain constant.
Since DBF is used to extend the communication range, the individual pre-BF SNR is low,  and the estimation errors  within the protocol can not be neglected and will lead to a combining phase error $\mpe{n}$. At the evaluation time $\mte$, the variance of the combining phase error $\mvarPe$ defined as $\mvar{\mpe{n}}$ is given by
$
	\label{eq:var_pe}
	\mvarPe=  (2\pi \mte)^2 \mvarF +  \mvarPh + \mvarFb 
$,
where the  frequency estimation  variance is given by $\mvarF=\mvar{\mfh{n}-\mfw{n}}$, the phase estimation and feedback variances are given by $\mvarPh=\mvar{\mph{n}-\mpwd{n}}$ and $\mvarFb=\mvar{\mpwd{n}-\mpw{n}}$ respectively. The value of these error variances depends on the estimators used.

\subsection{Estimators and Minimum Error Variance}
In this section, we describe the estimators used and their error variances. For frequency offset estimation, we calculate the phase difference between successive ZC repetitions according to the estimator in~\cite{yan_aeroconf_2019}, which has  an error variance
$
	\label{eq:freq_est_var}
	\mvarFe= \left(\frac{2 \mSNRms+(\mnzc-1)}{2 \mlzc(\mnzc-1)^2\mSNRms^{2}}\right)\frac{1}{(2\pi \mlzc \mTs)^{2}}
$~\cite[eq.70]{lank_semicoherent_1973}.
To reduce the estimation error, Kalman filter is applied to track the frequency error making the frequency error variance equal to 
$
	\label{eq:kf_op_var}
	\mvarF=\left(-q+q\sqrt{1+4\frac{\mvarFe}{q}}\right)/{2}
$
where $q$ is the process variance, which can be calculated from the oscillator datasheet~\cite{richard_brown_receiver-coordinated_2012}. The phase can be estimated by correlating with the known phase estimation preamble and then calculating arctan. The variance of this estimator is
$
	\label{eq:ph_est_var}
	\mvarPh=\frac{1}{2 \mnph \mSNRpre }
$~\cite{tretter_estimating_1985}.
As for the phase feedback, it is calculated by estimating the phase difference and its variance is given by
$
	\label{eq:fb_est_var}
	\mvarFb= \left(\frac{1}{\mnfb \mSNRms}+\frac{1}{2 \mnfb \mSNRms^{2}}\right)
$.
Substituting the previous equations in the definition of $\mvarPe$ yields an analytical expression for the combining phase error ($\mvarPe$) as a function of SNR and preamble lengths. This expression can be shown to be convex with respect to $\mnzc,\mnph,\mnfb$. From these estimators, we can see as expected that longer preambles reduce the estimation errors. However,  since the channel coherence time is limited, the BF overheads are assumed to upper  bounded by $\mnTrgt$. To attain the maximum coherent BF gain for a fixed overhead, we need to design the preambles to minimize the combining phase errors, which can be performed using
 \begin{align}
		& \underset{ \mnzc,\mnph,\mnfb }{\text{minimize}} & &  \mvarPe(\mnzc \mlzc,\mnph,\mnfb) \label{probII:obj} \\
	& \text{subject to} & &   \mnov(\mnzc \mlzc,\mnph,\mnfb)\leq  \mnTrgt 
	\nonumber\\ & & & 
	\mnzc,\mnph,\mnfb \in \Z^+, \ \ \mnzc\geq 2  \nonumber
\end{align}
This problem is a convex mixed integer problem and can be solved using CVX  with a mixed integer solver~\cite{grant_cvx_2014}.
 It can be argued that for this choice of estimators, the resulting combining phase error is zero mean Gaussian~\cite{hanna_distributed_2021}. We want to relate this phase error to the the BF gain.

\providecommand{\mGperc}{G_{10}}
\subsection{Combining Beamforming Gain Distribution}
Since the phase error is random, the resulting combining BF gain is also a random variable. Using their relation, we can derive the distribution of $G$
\newcommand{\mgammarv}{X_{\gamma}}
\begin{prop}
	\label{prop:g_gamma}
	Assuming a zero mean Gaussian combining phase error having variance $\mvarPe$, for small   $\mvarPe$ or for large $N$, the distribution of $G$ can be approximated by $N-\mgammarv$ where $\mgammarv$ is a random variable following the Gamma distribution $\mgammarv \sim \Gamma(K,\theta)$ with
$
		\label{eq:g_gamma_k}
		K=\frac{N(N-1)}{(1-e^{-\mvarPe})^{2}+2Ne^{-\mvarPe}}
$ and
$
		\label{eq:g_gamma_theta}
		\theta=\frac{1}{N} (1-e^{-\mvarPe})\left((1-e^{-\mvarPe})^{2}+2Ne^{-\mvarPe})\right)
$.
\end{prop}
This proposition is proved in~\cite{hanna_distributed_2021}.  The distribution of $G$ will be used to realize the required probability on the postBF SNR.%
\newcommand{\mFdist}{F}
\section{Calculating the Maximum Range}

To solve Problem (\ref{probI:obj})-(\ref{probI:probability}), we use a graphical approach. We first plot the achievable BF gain realizing the desired probability at each preBF SNR. Then we determine the points where the preBF SNR and the BF gain meet or exceed the required postBF SNR. Among them, we choose the one having the smallest preBF SNR, which corresponds to the largest distance.
First, we start by accounting  for the required probability by  rewriting (\ref{probI:probability})  in terms of the CDF of $G$ denoted by $\mFdist$ as follows $\mFdist \left(G=\mGreq \right)\leq 1-\mpmin$, where $\mGreq=\frac{\mSNRreq}{N \mSNRpre}$.
This can be rewritten as $G_{1-\mpmin}  \geq \mGreq$, where  $G_{1-\mpmin}=\mFdist^{-1}(1-\mpmin)$. The satisfaction of the last equation implies that the required post BF SNR is met at least  with the required probability. We are interested in the smallest $\mSNRpre$ corresponding to the largest $d$, which satisfies this relation.
 
 \begin{figure*}[t!]
 	\centering
 	\subfloat[Number of DBF radios ($N$).\label{fig:presnr_gain_N}]{\includegraphics[scale = 1]{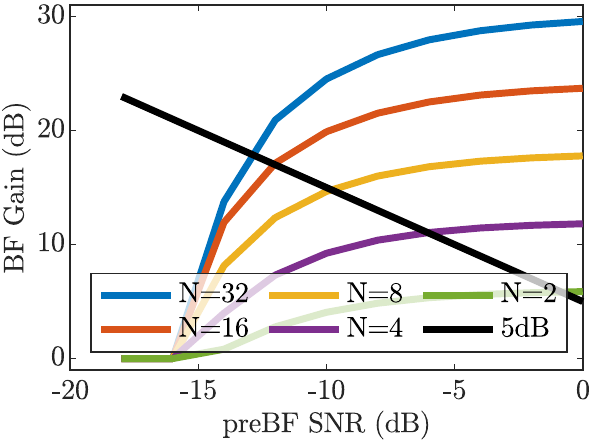}}
 	\subfloat[Destination Power ($\mdp$).\label{fig:presnr_gain_mastPwr}]{\includegraphics[scale = 1]{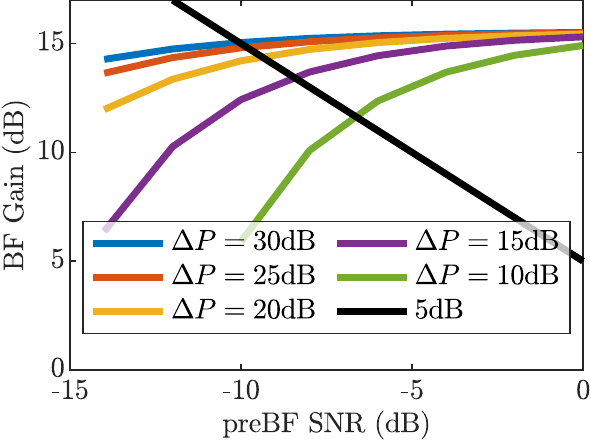}}
 	\subfloat[BF overhead length ($\mnTrgt$).\label{fig:presnr_gain_overhead}]{\includegraphics[scale = 1]{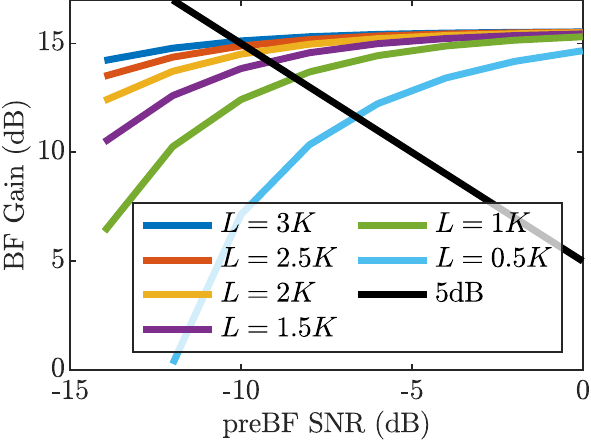}}
 	\caption{The relation of the pre-BF SNR and the BF gain as a function several parameters considering $\mSNRreq=5$dB with $\mpmin=0.9$. \label{fig:presnr_gain}}
 \end{figure*}
 \begin{figure}[t!]
 	\centerline{\includegraphics[scale = 1]{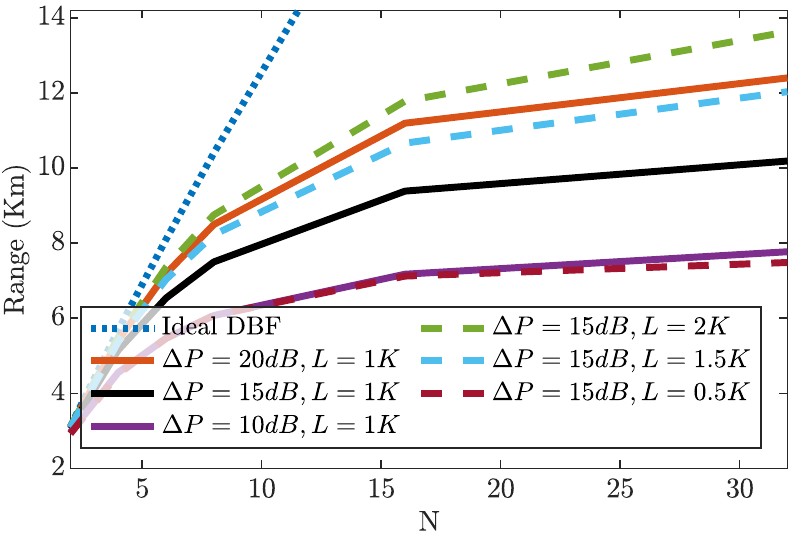}}
 	\caption{The max. communication range as a function of $N$.}
 	\label{fig:range_N}
 \end{figure}
 
Both $\mGreq$ and $G_{1-\mpmin}$  are  a function of $\mSNRpre$. By definition,  $\mGreq$ depends on $\mSNRpre$. As for $G_{1-\mpmin}$, since $\mFdist$ is the CDF of the Gamma random variable from Proposition~\ref{prop:g_gamma}, and its value depends on $N$ and $\mvarPe$, which in turn depends on the preBF SNR. Thus $G_{1-\mpmin}(\mSNRpre)$ is also a function of the preBF SNR. To solve the problem, we plot both $\mGreq(\mSNRpre)$ and $G_{1-\mpmin}(\mSNRpre)$ as a function of  $\mSNRpre$. Then we find the smallest $\mSNRpre$ satisfying the requirement.

\section{Results}
 Unless otherwise stated, we consider $N=6$ DBF radios, a maximum DBF overhead length of $\mnTrgt=1000$ samples, and that the destination SNR is higher by $\Delta P=15 dB$. The payload length is $N_p=8000$ and the guard times are chosen to make $t_e=5000 T_s$. The required SNR $\mSNRreq=5dB$ and must be satisfied by at least $\mpmin=0.9$.

To calculate the maximum range, we start by plotting the total BF gain ($N G$) in dB as a function of preBF SNR for different $N$ in Fig.~\ref{fig:presnr_gain_N}. Looking at the curve for $N=16$ in red as an example, each point in the plot is obtained by optimizing the length of each preamble within the BF overhead by solving Problem (\ref{probII:obj}). After optimizing and calculating $\mvarPe$ and using $N=6$, we know the CDF of the Gamma random variable using Proposition~\ref{prop:g_gamma}, which we use to calculate $G_{1-\mpmin}$, then $N G_{1-\mpmin}$ is plotted in dB. These steps are repeated for every point to yield the red curve for $N=16$. As expected, the smaller $\mSNRpre$, the smaller the BF gains, which drop to zero below a certain values. For large $\mSNRpre$, the coherent combining is almost perfect, and the curve starts to saturate with the whole $N^2$ BF gain. When plotted in dB, the constraint on the required BF gain ($N \mGreq$) becomes a straight line. The smaller the preBF SNR, the larger the required BF gain to meet the required SNR, and this shown for $\mSNRreq=5dB$ as the black line. All the points, where $N G_{1-\mpmin} \geq N \mGreq$ meet the required SNR, and the one corresponding to smallest $\mSNRpre$ is the intersection between the two lines.

In in Fig.~\ref{fig:presnr_gain_N}, we plotted $N G_{1-\mpmin}$ for different $N$. As $N$ increases, the maximum achievable BF gain ($N^2$) increases, which corresponds to the saturation value of the curves. The smallest $\mSNRpre$ meeting the requirement for different $N$  corresponds to the intersection between the $N G_{1-\mpmin}$ colored lines and  the $N \mGreq$ black line (which is equal to $\frac{\mSNRreq}{\mSNRpre}$ and does not depend on $N$). We can see that as $N$ increases the improvement in the minimum SNR starts saturating. This occurs because the total BF overheads are fixed. Increasing $N$ makes the duration of the preambles shorter, which results in larger phase errors and limits the improvement from larger $N$.  In  Fig.~\ref{fig:presnr_gain_mastPwr} and~\ref{fig:presnr_gain_overhead}, we repeat the same steps for $N=6$ and we increase the power of the destination ($\mdp$) and the overhead lengths respectively ($\mnTrgt$). First notice that in both Figures,  increasing either parameters has the same effect and only improves the BF gain  at low SNR. At high SNR, the estimation is almost perfect and since $N$ is fixed, the maximum BF gain can not exceed $N^2$ regardless of $\mdp$ or $\mnTrgt$. Increasing $\mdp$ improves the SNR of signals sent from the destination (synch preamble and channel feedback) reducing the estimation errors. As for increasing $L$, it enables increasing the preamble length and thus also reduces the estimation errors.

To translate the minimum SNR to deployment distance ($d$), we consider a channel exponent $k=2.3$, BF radios with $P_T=0$dBm with a 3dB noise figure, and a noise bandwidth of 1MHz. For different values of $N$, $\mnTrgt$, and $\mdp$, we calculate the maximum $d$, by calculating the minimum SNR from the intersection between $N G_{1-\mpmin}$ and $N \mGreq$  as previously discussed and then calculate the corresponding distance from the path loss. The results are shown in Fig.~\ref{fig:range_N} as a function of $N$ for different $\mnTrgt$ in dashed and $\mdp$ in solid. In dotted, we show the ideal communication range calculated by assuming the full $N^2$ BF gain was achieved.  First, notice that the attained range extension is well below  the ideal specially at high $N$ (corresponding to large distances and low preBF SNRs) because of the estimation errors. As we increase  $\mdp$ or $\mnTrgt$, these errors  decrease which can increase the communication range by up to 1.7x. Increasing $N$ as previously discussed increases the communication range, however, this increase has diminishing gains with $N$ because accommodating more signals within the limited BF overhead duration leads to even larger estimation errors. However, as we increase either $\mdp$ or $\mnTrgt$, the range improvement caused a larger $N$ increases. This shows that just increasing $N$ might not be a good strategy for extending the range as the improvement in BF gains diminishes due to estimation errors. However, using longer overheads or increased destination power is important to counter the diminishing gains as $N$ increases.

\section{Conclusion}
In this paper, for a destination-led DBF protocol, we calculated the maximum communication range realizing a desired post-BF SNR with a given probability. To calculate this maximum range, we analyzed and plotted the relation between the pre-BF SNR and BF gain, from which we determined the maximum distance meeting the requirements. The results show that at large distances, the synchronization and channel estimation errors dominate. In this regime, increasing the number of BF radios provides little improvement for the communication range. Reducing these errors by either increasing the duration of the preambles or increasing the destination power has a  more significant impact and can lead to larger range extension.
\clearpage
\pagebreak
\bibliographystyle{IEEEtran}
\bibliography{references}

\end{document}